\documentclass[11pt, reprint,prl,superscriptaddress,nofootinbib,floatfix,preprintnumbers]{revtex4-1}
\usepackage{graphicx, bm, color, amssymb, amsmath,subfigure}

\newcommand{\CP}{C\!P}

\begin{document}
\preprint{MAN/HEP/2015/14, TUM-HEP-1008/15, UMD-PP-015-012}
\title{Unified explanation of the $eejj$, diboson and dijet resonances at the LHC}

\author{P. S. Bhupal Dev}
\affiliation{Consortium for Fundamental Physics, School of Physics and Astronomy, University of Manchester, Manchester M13 9PL, United Kingdom}
\affiliation{Physik Department T30d,
Technische Universit\"{a}t M\"{u}nchen, James-Franck-Stra\ss e 1, 85748 Garching, Germany}

\author{R. N. Mohapatra}
\affiliation{Maryland Center for Fundamental Physics and Department of Physics, University of Maryland, College Park, Maryland 20742, USA}

\begin{abstract}
We show that the excess events observed in a number of recent LHC resonance searches can be simultaneously explained within a non-supersymmetric left-right inverse seesaw model for neutrino masses with $W_R$ mass around 1.9 TeV.  The minimal particle content that leads to gauge coupling unification in this model predicts $g_R\simeq 0.51$ at the TeV-scale, which is consistent with data. The extra color-singlet, $SU(2)$-triplet fermions required for unification can be interpreted as the Dark Matter of the Universe. Future measurements of the ratio of same-sign to opposite-sign dilepton events can provide a way to distinguish this scenario from the canonical cases of type-I and inverse seesaw, i.e. provide a measure of the relative magnitudes of the Dirac and Majorana masses of the right-handed neutrinos in the $SU(2)_R$-doublet of the left-right symmetric model.
\end{abstract}

\maketitle
\section{Introduction}\label{sec:1}
Recently, a number of resonance searches at the $\sqrt s=8$ TeV LHC have reported a handful of excess events around invariant mass of 1.8 -- 2 TeV. The most significant ones are: (i) a $3.4\sigma$  local ($2.5\sigma$ global) excess in the ATLAS search~\cite{ATLAS1} (see also~\cite{CMS-VV} for the corresponding CMS searches reporting a mild excess at the same mass) for a heavy resonance decaying into a pair of Standard Model (SM) gauge bosons $VV$ (with $V=W,Z$); (ii) a $2.8\sigma$ excess in the CMS search~\cite{CMS}  for a heavy right-handed (RH) gauge boson $W_R$ decaying  into an electron and RH neutrino $N_R$, whose further decay gives an $eejj$ final state; (iii) a $2.2\sigma$ excess in the CMS search~\cite{CMS1} for $W'\to WH$, where the SM Higgs boson $H$ decays into $b\bar{b}$ and $W\to \ell \nu$ (with $\ell=e,\mu$); (iv) a $2.1\sigma$ excess in the CMS dijet search~\cite{CMS2}. These excesses of course need to be confirmed with more statistics at the LHC run II before any firm conclusion about their origin can be derived. Nevertheless, taking them as possible indications of new physics beyond the SM, it is worthwhile to examine whether all of them could be simultaneously explained within a self-consistent, ultra-violet complete theory that could be tested in foreseeable future. 

One class of models that seems to broadly fit the observed features in all the above-mentioned channels is the Left-Right Symmetric Model (LRSM) of weak interactions based on the gauge group $SU(2)_L\times SU(2)_R\times U(1)_{B-L}$~\cite{LR}, with the RH charged gauge boson mass $M_{W_R}\sim 2$ TeV and with $g_R<g_L$ at the TeV-scale~\cite{Dobrescu:2015qna}, $g_{L(R)}$ being the $SU(2)_{L(R)}$ gauge coupling strength.  Within this framework, the $eejj$ excess~\cite{CMS} can be understood~\cite{Dobrescu:2015qna, Deppisch:2014qpa, Fowlie:2014iua, Gluza:2015goa} as $pp\to W_R\to eN_R\to eejj$~\cite{KS} and is related to the type-I seesaw mechanism~\cite{seesaw} for neutrino masses. The $WZ$ excess~\cite{ATLAS1} and $WH$ excess~\cite{CMS1} can be understood~\cite{Dobrescu:2015qna, Gao:2015irw, Brehmer:2015cia} in terms of $W_R\to WZ,~WH$, since these couplings naturally arise in these models from the vacuum expectation values (VEVs) of the bidoublet field used to generate quark and lepton masses~\cite{LR} (see~\cite{other} for some alternative explanations of the diboson excess). Finally, the dijet excess~\cite{CMS2} can simply be due to $W_R\to jj$.

However, a particular aspect of the observations in the $eejj$ channel, namely,  a suppression of same-sign electron pairs with respect to opposite-sign pairs~\cite{CMS}, cannot be explained within the minimal LRSM with type-I seesaw mechanism. This is because of the fact that for a type-I seesaw interpretation of the $eejj$ excess, one expects equal number of same and opposite-sign dileptons due to the purely Majorana nature of the RH neutrinos (see~\cite{Gluza:2015goa} for an exception, when the interference of two non-degenerate RH Majorana neutrinos with mixed flavor content
and opposite $\CP$ parities can partially suppress the same-sign dilepton signal). Thus, a heavy pseudo-Dirac neutrino, as naturally occurs in the inverse seesaw mechanism~\cite{inverse}, seems to be the simplest possibility to explain the suppression of same-sign dilepton events in both CMS~\cite{CMS, Khachatryan:2015gha} and ATLAS~\cite{ATLAS2} searches.

The main result of this letter is that if the difference between same and opposite-sign dilepton signal becomes statistically significant, it more likely suggests an inverse seesaw interpretation rather than a type-I seesaw in a left-right (LR) model. Note that in the original inverse seesaw proposal~\cite{inverse}, the lepton number violation is small, being directly proportional to the light neutrino masses, and hence, it is rather unlikely to have {\em any} same-sign dilepton events in this scenario. In this letter, we show that there exists a class of inverse seesaw models where the heavy neutrinos are still Majorana fermions with non-negligible lepton number violation, without affecting the inverse seesaw neutrino mass formula.  In this class of models, it is possible to accommodate a {\em non-zero} same-sign dilepton signal, while being consistent with the suppression with respect to the opposite-sign signal. In particular, a statistically significant non-zero ratio of same and opposite-sign dilepton signal events could be used to test the relative strength between the Dirac and Majorana nature of the heavy neutrinos at the LHC.

Another important result of this letter is that our TeV-scale LRSM with inverse seesaw unifies to an  $SO(10)$ Grand Unified Theory (GUT) at a high scale $M_U\sim 10^{17}$ GeV without introducing any other intermediate scales, which is remarkable for a non-supersymmetric theory. This is achieved with a minimal TeV-scale particle content, which predicts the value of $g_R\simeq 0.51$ at the TeV-scale, thus naturally satisfying the requirement $g_R<g_L$ to explain the excess events mentioned above. Moreover, such a single-step unification without introducing supersymmetry (SUSY) also requires the existence of $SU(2)$-triplet fermions, which can naturally act as the Dark Matter of the Universe~\cite{HP}. Finally, this model could also explain the observed baryon asymmetry of the Universe via leptogenesis through the out-of-equilibrium decay of the heavy Majorana neutrinos, while avoiding the stringent leptogenesis bounds on $M_{W_R}$~\cite{DLM} due to suppressed $W_R$-induced washout in this case~\cite{BDM}.


\section{The Model}
As in the usual LRSM~\cite{LR}, denoting
$Q\equiv (u ~ d)^{\sf T}$ and $\psi\equiv (\nu_\ell ~ \ell)^{\sf T}$ as the quark and lepton doublets respectively, $Q_{L}$ and $\psi_{L}$ are doublets under $SU(2)_{L}$,
 while $Q_R$ and $\psi_R$  are $SU(2)_R$ doublets. 
To implement the inverse seesaw mechanism~\cite{inverse}, we add three gauge-singlet fermions $S_i$.  
The minimal Higgs sector of the model consists of an $SU(2)_L\times SU(2)_R$ bidoublet $\phi$, an $SU(2)_R$-doublet $\chi_R$ and $SU(2)_R$-triplet $\Delta_R$.
The gauge symmetry $SU(2)_R\times U(1)_{B-L}$ is broken to $U(1)_Y$ by the triplet VEV $\langle \Delta^0_R\rangle = v_R$, as well as the doublet VEV $\langle \chi^0_R\rangle =\kappa_R$, whereas the bidoublet VEV $\langle\phi\rangle={\rm diag}(\kappa, \kappa')$ breaks the SM gauge group $SU(2)_L\times U(1)_Y$ to $U(1)_{\rm em}$.
The LH counterparts ($\chi_L$ and $\Delta_L$) are assumed to be at the GUT scale, following ``broken $D$-parity" models~\cite{CMP}.

The Yukawa Lagrangian of the model is given by
\begin{align}
{\cal L}_Y  = \ & h^{q}_{ij}\bar{Q}_{L,i}\phi Q_{R,j}+\tilde{h}^{q}_{ij}\bar{Q}_{L,i}\tilde{\phi} Q_{R,j}+
h^{l}_{ij}\bar{\psi}_{L,i}\phi \psi_{R,j} \nonumber \\
& + \tilde{h}^{l}_{ij}\bar{\psi}_{L,i}\tilde{\phi}\psi_{R,j}+f_{ij}\bar{\psi}_{R,i} \chi_R S_j +\frac{1}{2}S^C_i \mu_{S,ij} S_j
 \nonumber \\
& +\frac{1}{2}f^\prime_{ij} (\psi^C_{R,i}\Delta_R \psi_{R,j} +\psi^C_{L,i}\Delta_L \psi_{L,j})+ {\rm H.c.},
\label{eq:yuk}
\end{align}
where $i,j=1,2,3$ stand for the fermion generations, $C$ for charge conjugation and $\tilde{\phi}=\tau_2\phi^*\tau_2$ ($\tau_2$ being the second Pauli matrix). 
After electroweak symmetry breaking, the Dirac fermion masses are given by the generic formula $M_f~=~h^f\kappa + \tilde{h}^f\kappa'$ for up-type fermions, while for down-type quarks and
charged-leptons, it is the same formula with $\kappa\leftrightarrow \kappa'$.  Eq.~(\ref{eq:yuk}) leads to the Dirac mass matrix  for neutrinos: $M_D = h^{l}\kappa + \tilde{h^{l}}\kappa'$.  The RH neutrinos $N_R$ and their singlet partners $S$ generate a Dirac mass term $M_N=f\kappa_R$, while the RH neutrinos also get a Majorana mass $\mu_R=f'v_R$. Thus, the generalized inverse seesaw  neutrino mass matrix in the flavor basis $\{\nu^C,N,S^C\}$ is given by~\cite{gavela, DP12, Parida:2012sq, Abada:2014vea}
\begin{eqnarray}
{\cal M}_\nu \ = \ \left(\begin{array}{ccc} 0 & M_D&0\\M^{\sf T}_D & \mu_R & M_N^{\sf T}\\
0 & M_N & \mu_S\end{array}\right) \; .
\label{eq:numass}
\end{eqnarray}
At tree-level, the Majorana entry $\mu_R$ in Eq.~\eqref{eq:numass} does not affect the light neutrino masses, which are only proportional to $\mu_S$. However, standard electroweak radiative corrections~\cite{AP92} give an one-loop contribution proportional to $f_N\mu_R$, where the loop-factor is given by~\cite{DP12}
\begin{align}
f_N =  \frac{\alpha_{W}}{16\pi}\left[\frac{M_H^2}{M_N^2-M_H^2}\ln\left(\frac{M_N^2}{M_H^2}\right)+ \frac{3M_Z^2}{M_N^2-M_Z^2}\ln\left(\frac{M_N^2}{M_Z^2}\right) \right]
\end{align}
with $\alpha_W=g_L^2/4\pi$. Combining the tree and loop contributions, we can write an effective light neutrino Majorana mass which, in the one-generation case, becomes
\begin{eqnarray}
M_{\nu} \ \simeq \ \left(\mu_S+f_N\mu_R\frac{M_N^2}{M_W^2}\right)\left(\frac{M_D}{M_N}\right)^2 \; .
\label{nu1}
\end{eqnarray}
More importantly, the other two mass eigenstates which are linear combinations of $N$ and $S$ are also Majorana fermions. Denoting them by $({N}_1, N_2)$, we get the corresponding masses from Eq.~\eqref{eq:numass}:
\begin{eqnarray}
M_{N_{1,2}} \ \simeq \ \frac{1}{2}\left[\mu_R\pm \sqrt{\mu_R^2+4M_N^2}\right] \; ,
\label{nu2}
\end{eqnarray}
where we have assumed that $\mu_S \ll \mu_R$ and $M_D\ll M_N$, as suggested by Eq.~\eqref{nu1}.
From Eq.~\eqref{nu2}, we see that for $\mu_R \ll M_N$, $N_{1,2}$ form a pseudo-Dirac pair, as in the usual inverse seesaw case. In the other extreme limit  $\mu_R \gg M_N$, $N_1$ is purely Majorana with $M_{N_1}=\mu_R$ as in the type-I seesaw, whereas $N_2$ will have a mass  proportional to $\mu_S$. Thus, for intermediate values of $\mu_R$, we can have scenarios with varying degree of lepton number breaking. This has important consequences for lepton number violating signals in colliders (see below), as well as in neutrinoless double beta decay experiments~\cite{Parida:2012sq}. 

\section{Lepton Number Violating Signal}
The charged-current Lagrangian in the RH sector is given by
\begin{align}
-{\cal L}_{\rm CC}^R \ = \ & \frac{g_R}{2\sqrt 2}W_R^{\mu-}\bar{\ell}_R\gamma_\mu (1+\gamma_5) U_{\ell N}[\cos\theta N_1 - \sin\theta N_2] \nonumber \\
& \qquad \qquad +{\rm H.c.},
\label{eq:CC}
\end{align}
where $U_{\ell N}$ denotes the mixing between the mass eigenstates $N_i$ and the flavor eigenstate $N_\ell$, and $\theta$ is the mixing angle between $N_{1,2}$ with $\tan{2\theta} = 2 M_N/\mu_R$.

For dilepton final states, we observe from Eq.~\eqref{eq:CC} that for $M_{N_{1,2}}<M_{W_R}$, the on-shell production of $W_R$ in the $pp$ collisions at the LHC is followed by its decay to an admixture of the mass eigenstates $N_{1,2}$ and their subsequent decay to $\ell jj$. Thus, for $M_{N_1}\neq M_{N_2}$ for the $N_\ell$ flavor, we do not expect to see a single peak in the $\ell jj$ invariant mass distribution, which could naturally explain the absence of such a peak in the CMS data~\cite{CMS}.

Also for $M_N\neq 0$, the heavy neutrino mass eigenstates $N_{1,2}$ in Eq.~\eqref{nu2} have masses of opposite sign, which means that one of them is odd under $\CP$ transformation~\cite{LW}, while the other is even under $\CP$. Hence, there will be a relative sign between the amplitudes of the processes for opposite-sign ($pp\to W_R \to \ell^\pm N_{1,2} \to \ell^\pm \ell^\mp jj$) and same-sign ($pp\to W_R \to \ell^\pm N_{1,2} \to \ell^\pm \ell^\pm jj$) dilepton final states. The ratio of the amplitudes is given by
\begin{eqnarray}
r \ \equiv \ \frac{{\cal A}_{\ell^+\ell^+jj}}{{\cal A}_{\ell^+\ell^-jj}} \ \simeq \ \cos {2\theta} \equiv \sqrt{\frac{\mu_R^2}{\mu_R^2+4M_N^2}} \; ,
\label{ratio}
\end{eqnarray}
where we have used the field theoretic method of~\cite{Grimus:1998uh} to account for the coherent mixing of the two propagating mass eigenstates $N_{1,2}$.
Eq.~\eqref{ratio} implies that unlike the type-I seesaw case, the number of same-sign and opposite-sign dilepton events can be different in our case and the ratio $r^2=\cos^2{2\theta}$, which can be anywhere between 0 and 1, will be a measure of the relative magnitude of the two entries $\mu_R$ (Majorana) and $M_N$ (Dirac) in the inverse seesaw mass matrix~\eqref{eq:numass}. This can therefore provide a plausible mechanism for understanding a non-zero but smaller number of same-sign dilepton events than the opposite-sign dilepton events, as observed by the CMS (barring statistical fluctuations).

In a simple model, one can assume both the Dirac and Majorana masses of the $(N,S)$ sector of the heavy neutrino mass matrices to be flavor diagonal. In this case, the ``small" Majorana entry $\mu_S$ can give the desired flavor structure to satisfy the neutrino oscillation data, without affecting the dilepton signal, as long as $M_D\lesssim M_W$ [cf. Eq.~\eqref{nu1}]. Also for final states with different flavors, the ratio $r$ in Eq.~\eqref{ratio} can be different, depending on the structure of the full neutrino mass matrix~\eqref{eq:numass} in the 3-generation case. Since we are mainly interested in explaining the $eejj$ excess, we will not give a detailed description of the possible flavor structures. Example fits to neutrino data can be found in~\cite{DM, DP12}.

It is also worth noticing that the same characteristic of dilepton charge content as in Eq.~\eqref{ratio} could emerge in the context of SM inverse seesaw without the LR symmetry, provided the mixing between light and heavy neutrinos (typically denoted by $V_{\ell N}\sim M_DM_N^{-1}$ in the literature) is sizable, so that the heavy neutrinos could be produced on-shell with an observable cross section through the Drell-Yan process mediated by an $s$-channel $W$ boson~\cite{Deppisch:2015qwa}. However, the invariant mass pattern of the final state jets will be different, since in this case they will come from an on-shell $W$ decay, and not from an off-shell $W_R$ decay as in the LR case. One could also have a dominant 2-body decay $N_R\to W\ell^\pm$ in the LR case, if the mixing parameter $V_{\ell N}$ is large~\cite{CDM}, but here we do not include this possibility, simply because this effect could be masqueraded by the hitherto unknown mixing parameter $U_{\ell N}$ in the RH neutrino sector.

\section{Gauge coupling unification}
Now let us see whether our inverse seesaw LR model with TeV-scale $W_R$ can emerge from a non-SUSY $SO(10)$ GUT, without introducing any intermediate scales, in the same spirit as the SUSY version~\cite{DM}. We find it is possible, if the particle content given above is supplemented by two Weyl fermion triplets $\Sigma_L({\bf 3}, {\bf 1},0,{\bf 1})\oplus \Sigma_R({\bf 1},{\bf 3},0,{\bf 1})$ and an extra real scalar multiplet $\varphi ({\bf 1},{\bf 3},0,{\bf 8})$ under $SU(2)_L\times SU(2)_R\times U(1)_{B-L}\times SU(3)_c$, all with masses in the TeV scale. The triplet fermions $\Sigma_{L,R}$ could be assumed to have come from a {\bf 45}-dimensional fermion multiplet of $SO(10)$, with its remaining components at the GUT scale. 
 Similarly the color-octet scalar $\varphi$ could originate from a {\bf 210}-dimensional $SO(10)$ scalar multiplet which is often used to break the $SO(10)$ gauge symmetry.

\begin{figure}[t]
\centering
\includegraphics[width=7cm]{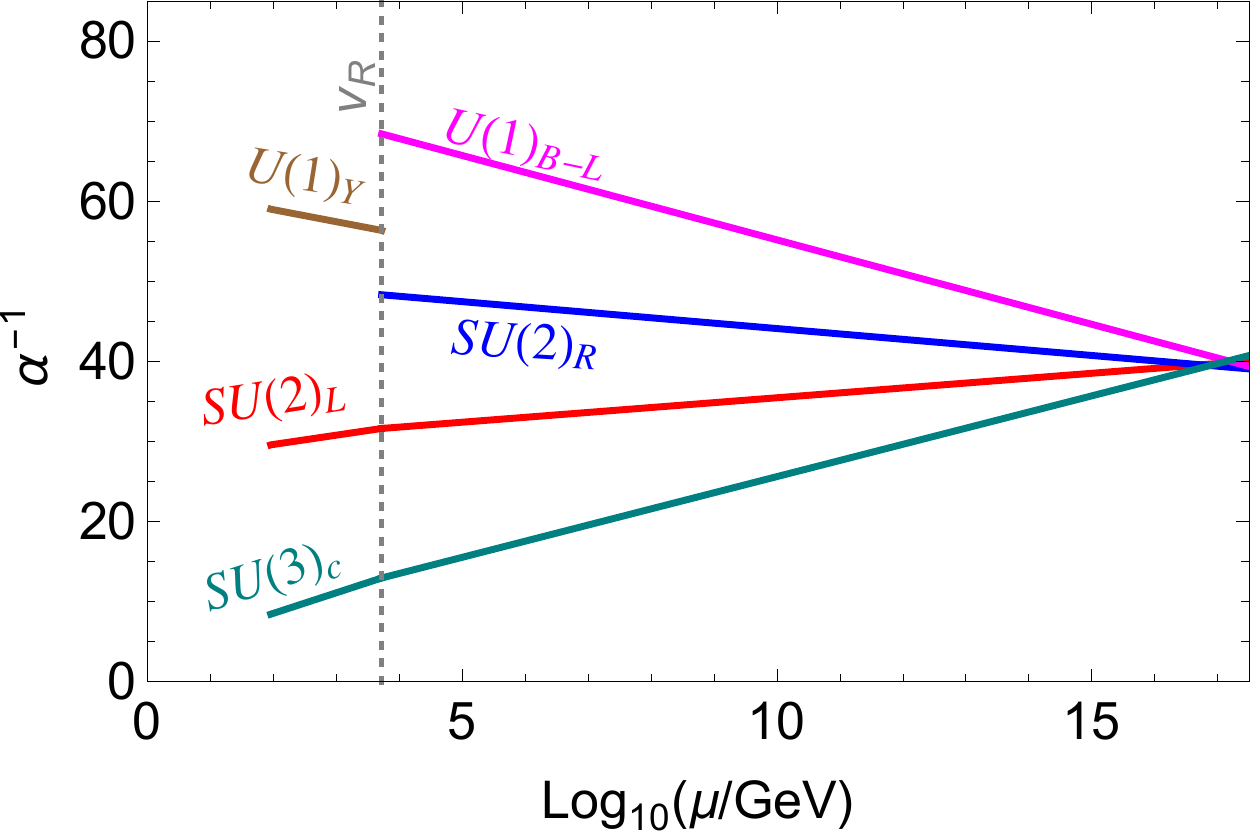}
\caption{Gauge coupling unification in our inverse seesaw Left-Right model with $M_{W_R}=1.9$ TeV. }
\label{fig1}
\end{figure}

The evolution of the gauge couplings is shown in Figure~\ref{fig1}, where we have chosen the $SU(2)_R$ symmetry-breaking scale such that $M_{W_R}=1.9$ TeV, as preferred  by the various excesses alluded to above (see next section). In Figure~\ref{fig1}, we have evolved the gauge couplings $\alpha_i\equiv g_i^2/4\pi$ from $M_Z$ to $v_R$ by the SM one-loop $\beta$-functions $(b_{2L},b_{1Y},b_{3c})=\left( -\frac{19}{6},\frac{41}{10},-7 \right)$. At $v_R$, the symmetry expands to $SU(2)_L\times SU(2)_R\times U(1)_{B-L}\times SU(3)_c$ with all the SM fermions transforming as LR multiplets together with the new postulated fields above. The one-loop $\beta$-functions in this case are given by $(b_{2L},b_{2R},b_{B-L},b_{3c})=
\left(-\frac{5}{3}, \frac{11}{6}, \frac{23}{4}, -\frac{11}{2}\right)$. The couplings unify around $10^{17}$ GeV, which easily satisfies the proton decay constraints.

We note some salient features of this model: 

(i) $\beta_{2L}\neq \beta_{2R}$ which implies $g_L(v_R)\neq g_R(v_R)$. This can be implemented by breaking $D$-parity at a higher scale~\cite{CMP}.

(ii) Unification with this particle content predicts $g_R(v_R)\simeq 0.51 < g_L(v_R)$. This is still consistent with what is required to fit all the observed excesses (see next section). Note that in our case, there is a lower limit on $g_R$ set by the matching condition $\alpha_{1Y}^{-1}(v_R) = \frac{3}{5}\alpha_{2R}^{-1}(v_R)+\frac{2}{5}\alpha_{B-L}^{-1}(v_R)$, and therefore, it is not possible to choose $g_R/g_L \lesssim 0.76$, unlike in earlier analyses~\cite{Dobrescu:2015qna, Deppisch:2014qpa, Fowlie:2014iua, Gao:2015irw, Brehmer:2015cia}.

(iii) A detailed fermion mass and mixing fit can be done the same way as in~\cite{DM}, since the features relevant to fermion masses are similar in both models. Also note that our model is consistent with the low energy constraints coming from $K_L-K_S$ and $B-\overline{B}$ systems~\cite{fcnc}, which roughly imply $\left(\frac{g_R}{g_L}\right)^2\left(\frac{2.4~{\rm TeV}}{M_{W_R}}\right)^2<1$. Moreover, the $Z'$ boson mass predicted by these models is expected to be larger than $M_{W_R}$, thus making it consistent with the current LHC bounds on $Z'$ mass~\cite{Patra:2015bga}. Moreover, the $W_R$ contributions to lepton flavor violating (LFV) processes like $\mu\to e\gamma$ and $\mu\to e$ conversion rates, as well as the muon $(g-2)$ are well within the current limits for small $U_{\mu N}$, which is anyway desirable in our model to explain the absence of an excess in the $\mu\mu jj$ and $e\mu jj$  channels in the CMS search~\cite{CMS}. A detailed investigation of the model predictions for these LFV observables will be given elsewhere. 

(iv) The RH fermion triplet $\Sigma_R$ of the model may act as the Dark Matter of the Universe~\cite{HP} provided there is an additional $Z_2$ symmetry under which only the fermion triplet is odd, thus forbidding its coupling to the SM fermions and thereby making it stable. The LH component will rapidly self-annihilate due to its couplings to the SM gauge and Higgs bosons and will not affect the evolution of early Universe.

(v) The observed baryon asymmetry of the Universe could be explained via leptogenesis due to the out-of-equilibrium decay of the heavy Majorana neutrinos. Unlike the type-I seesaw case with $g_L=g_R$~\cite{DLM}, the $W_R$-induced washout processes in this model are expected to be under control~\cite{BDM} for $M_{W_R}$ around 2 TeV. A detailed quantitative analysis of the lepton asymmetry solving the relevant Boltzmann equations will be postponed to a future work.

\section{LHC Phenomenology}
Finally, we explore the viable model parameter space which can simultaneously fit all the observed excesses in $eejj$, $WZ$, $WH$ and $jj$ resonance searches. We calculate the $W_R$-production cross section at NLO using the FeynRules~\cite{Alloul:2013bka} implementation of the LRSM~\cite{Roitgrund:2014zka} in MadGraph~\cite{Alwall:2014hca} with NNPDF2.3 PDF sets~\cite{Ball:2012cx}. We find $\sigma(pp\to W_R)=223.4$ fb for $M_{W_R}=1.9$ TeV and $g_R=0.51$.  The relevant decay widths of $W_R$ are given in the Appendix. 

\begin{figure}[t!]
\centering
\includegraphics[width=8cm]{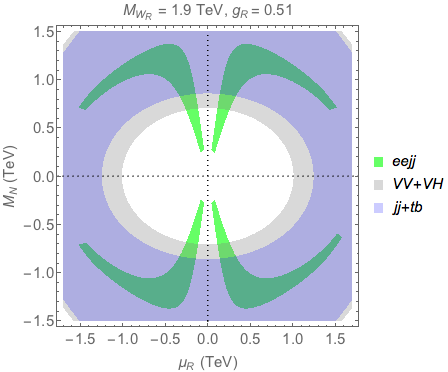}
\caption{The shaded regions represent the allowed $(\mu_R,M_N)$ parameter space. 
Note that the grey and blue shaded regions almost overlap. The intersection of the shaded blue and green regions simultaneously fits all the excesses.}
\label{fig2}
\end{figure}

For the $eejj$ excess, the total cross section can be fitted well with $U^2_{eN}\frac{g_R}{g_L}=0.3-0.5$~\cite{Deppisch:2014qpa}. 
In our model, $g_R/g_L=0.81$ at TeV-scale fixed by unification requirements, but we have the freedom in $U_{e N}$ without much restriction from neutrino oscillation data due to the inverse seesaw structure in Eq.~\eqref{eq:numass}. This also allows us to choose a small $U_{\mu N}$ to explain the absence of $\mu\mu jj$ and $e\mu jj$  excesses, as well as to satisfy the LFV constraints in the $e-\mu$ sector. The most important aspect of the $eejj$ excess in our model is explaining the observed ratio of 1/13 for the same-sign versus opposite-sign dilepton events~\cite{CMS}. From Eq.~\eqref{ratio}, we derive the allowed parameter space in the $\mu_R-M_N$ plane satisfying this constraint,  as shown in Figure~\ref{fig2} (green shaded regions). We have also demanded that in this allowed region, both the heavy neutrino masses $|M_{N_{1,2}}|$ contributing to the $eejj$ signal are above $0.1M_{W_R}$ and at least 100 GeV smaller than $M_{W_R}$, otherwise the signal efficiency after satisfying the CMS selection cuts~\cite{CMS} drops substantially and cannot explain the excess with the given production cross section.

For explaining the diboson and dijet excesses, we use the signal cross section values from a recent
fit~\cite{Brehmer:2015cia}: $\sigma_{WZ} = 5.9^{+5.3}_{-3.5}$ fb, $\sigma_{WH} = 4.5^{+6.2}_{-3.9}$ fb,
$\sigma_{jj} = 91^{+53}_{-45}$ fb and $\sigma_{tb} = 0^{+11}_{-0}$ fb, together with the selection efficiencies quoted in the corresponding experimental analyses. Note that since we have the additional decay modes $W_R\to N\ell$, as compared to the analysis in~\cite{Brehmer:2015cia} where $M_N>M_{W_R}$ was assumed to forbid this channel, all the $W_R$ branching ratios in our case will depend on the heavy neutrino masses. In particular, since the branching ratios to the diboson and dijet channels are somewhat reduced, it is now possible to fit the excesses in these channels with a relatively larger $g_R$ value, as dictated by unification in our case. The preferred regions in the $\mu_R-M_N$ plane satisfying the diboson (shaded gray) and dijet (shaded blue) excesses are shown in Figure~\ref{fig2}. It is clear that there exists a large parameter space simultaneously fitting all the excesses, while being consistent with gauge coupling unification, as well as other low-energy constraints. The only mild fine-tuning needed is to make $0<|\mu_R| < 2|M_N|$. 

\section{Conclusion}
We have presented a simple ultra-violet complete non-supersymmetric left-right model with inverse seesaw mechanism for neutrino masses that seems to fit all LHC excesses near 2 TeV invariant mass. 
Our model with a minimal TeV-scale particle content leads to successful gauge coupling unification at an experimentally allowed high scale, and predicts the value of the RH gauge coupling at the low-scale. This is shown to be consistent with our fit.  
Another main result of this letter is that the ratio of the dilepton signals with lepton number violating and conserving final states could be used to distinguish this class of inverse seesaw models with varying degree of lepton number violation from the canonical type-I (purely Majorana) and inverse seesaw (pseudo-Dirac) scenarios.  

\acknowledgements
\section{Acknowledgments}
We are grateful to Mrinal Dasgupta, Julian Heeck, Zhen Liu, Natsumi Nagata and Sudhanwa Patra for helpful discussions and comments. We also acknowledge  the Fermilab Theory group and the Mainz Institute for Theoretical Physics (MITP) for hospitality and partial support during the completion of this work. The work of P.S.B.D. is supported by the Lancaster-Manchester-Sheffield  Consortium  for  Fundamental  Physics under  STFC   grant  ST/L000520/1, as well as by a TUM University Foundation Fellowship and the DFG cluster of excellence ``Origin and Structure of the Universe". The work of R.N.M. is supported in part by the National Science Foundation Grant No. PHY-1315155.

\onecolumngrid
\appendix
\section{Appendix}
Here we list the analytic formulae for relevant partial decay widths of $W_R$:
\begin{align}
\Gamma(W^\pm_R\to N_i\ell^\pm)  \ = \ & \frac{g_R^2}{48\pi}|U'_{\ell N_i}|^2M_{W_R}\left(1+\frac{M_{N_i}^2}{2M_{W_R}^2}\right)\left(1-\frac{M_{N_i}^2}{M_{W_R}^2}\right)^2 \; , \\
\Gamma(W^+_R\to t\bar{b})  \ = \ & \frac{g_R^2}{16\pi}M_{W_R}\left(1+\frac{M_t^2}{2M_{W_R}^2}\right)\left(1-\frac{M_t^2}{M_{W_R}^2}\right)^2 \; , \\
\Gamma(W_R^+ \to u\bar{d}) \ = \ & \Gamma(W_R^+ \to c\bar{s}) \ = \ \frac{g_R^2}{16\pi}M_{W_R} \; , \\
\Gamma(W^\pm_R\to W^\pm Z)  \ = \ & \frac{g_R^2}{192\pi}M_{W_R}\sin^2 2\beta \left(1-2\frac{M_W^2+M_Z^2}{M_{W_R}^2}+\frac{(M_W^2-M_Z^2)^2}{M_{W_R}^4}\right)^{3/2}\nonumber \\
&\times  \left(1+10 \frac{M_W^2+M_Z^2}{M_{W_R}^2}+\frac{M_W^4+10M_W^2M_Z^2+M_Z^4}{M_{W_R}^4}\right) \; , \\
\Gamma(W^\pm_R\to W^\pm H)  \ = \ & \frac{g_R^2}{192\pi}M_{W_R}\cos^2 (\alpha+\beta) \left(1-2\frac{M_W^2+M_H^2}{M_{W_R}^2}+\frac{(M_W^2-M_H^2)^2}{M_{W_R}^4}\right)^{1/2}\nonumber \\
&\times  \left(1+\frac{10M_W^2-2M_H^2}{M_{W_R}^2}+\frac{(M_W^2-M_H^2)^2}{M_{W_R}^4}\right) \; ,
\end{align}
where $\theta_W$ is the weak mixing angle, $\tan \beta=\kappa_1/\kappa_2$ is the ratio of the bidoublet VEVs which characterizes the $W_L-W_R$ mixing $\xi \simeq -2\frac{g_R}{g_L}\frac{\tan\beta}{1+\tan^2\beta}\left(\frac{M_W}{M_{W_R}}\right)^2$ in $W_R \to WZ, WH$ decays, $\alpha$ is the mixing of the SM Higgs boson with the second Higgs doublet coming from the bidoublet in the model, and $U'_{\ell N_i}$ is the flavor mixing $U_{\ell N}$ in the RH neutrino sector multiplied by either $\cos^2\theta$ or $\sin^2\theta$ [cf. Eq.~(6) in the main text]. We assume that the non-standard Higgs bosons in our model do not contribute significantly to the total decay width of $W_R$. Note that in the SM Higgs alignment limit $\alpha\to \beta-\pi/2$,  the $W_R\to WH$ and $W_R\to WZ$ partial decay widths are equal (up to higher order terms in the various mass ratios), as is required by the Goldstone boson equivalence theorem. For our numerical fit, we have chosen the benchmark values $M_{W_R}=1.9$ TeV, $g_R=0.51$, $\beta=\pi/4$ and $\alpha=\beta-\pi/2$. This gives $|\xi|\sim 1.4\times 10^{-3}$, consistent with the experimental constraints~\cite{Langacker:1989xa}. 


\begin{thebibliography}{99}



\bibitem{ATLAS1} G. Aad {\em et al.} [ATLAS Collaboration],
arXiv:1506.00962 [hep-ex]. See also G.~Aad {\it et al.} [ATLAS Collaboration],
  Eur.\ Phys.\ J.\ C {\bf 75}, 69 (2015) 
  [arXiv:1409.6190 [hep-ex]]; 
  Eur.\ Phys.\ J.\ C {\bf 75}, 209 (2015)
  [arXiv:1503.04677 [hep-ex]].

\bibitem{CMS-VV} 
V.~Khachatryan {\it et al.} [CMS Collaboration],
  JHEP {\bf 1408}, 173 (2014)
  [arXiv:1405.1994 [hep-ex]]; 
  JHEP {\bf 1408}, 174 (2014)
  [arXiv:1405.3447 [hep-ex]]; 


\bibitem{CMS} V. Khachatryan {\em et al.} [CMS Collaboration],
Eur. Phys. J. {\bf C 74}, 3149 (2014) [arXiv:1407.3683 [hep-ex]].

\bibitem{CMS1} CMS Collaboration,
CMS-PAS-EXO-14-010 (2015). 

\bibitem{CMS2} V. Khachatryan {\em et al.} [CMS Collaboration],
Phys. Rev. D {\bf 91}, 052009 (2015) [arXiv:1501.04198 [hep-ex]]. See also 
G.~Aad {\it et al.} [ATLAS Collaboration],
  Phys.\ Rev.\ D {\bf 91}, 052007 (2015)
  [arXiv:1407.1376 [hep-ex]].

\bibitem{LR} J. C. Pati and A. Salam, Phys. Rev. D {\bf 10}, 275 (1974);
R. N. Mohapatra and J. C. Pati, Phys. Rev. D {\bf 11}, 566 (1975);
Phys. Rev. D {\bf 11},  2558 (1975);
G. Senjanovi\'{c} and R. N. Mohapatra, Phys. Rev. D {\bf 12} 1502 (1975).

\bibitem{Dobrescu:2015qna}  B.~A.~Dobrescu and Z.~Liu,
  arXiv:1506.06736 [hep-ph];
  arXiv:1507.01923 [hep-ph].


\bibitem{Deppisch:2014qpa} F.~F.~Deppisch, T.~E.~Gonzalo, S.~Patra, N.~Sahu and U.~Sarkar,
  Phys.\ Rev.\ D {\bf 90}, 053014 (2014) [arXiv:1407.5384 [hep-ph]];
Phys.\ Rev.\ D {\bf 91}, 015018 (2015)
  [arXiv:1410.6427 [hep-ph]].

\bibitem{Fowlie:2014iua}
M.~Heikinheimo, M.~Raidal and C.~Spethmann,
  Eur.\ Phys.\ J.\ C {\bf 74}, 3107 (2014)
  [arXiv:1407.6908 [hep-ph]];
J.~A.~Aguilar-Saavedra and F.~R.~Joaquim,
  Phys.\ Rev.\ D {\bf 90}, 115010 (2014)
  [arXiv:1408.2456 [hep-ph]];
A.~Fowlie and L.~Marzola,
  Nucl.\ Phys.\ B {\bf 889}, 36 (2014) [arXiv:1408.6699 [hep-ph]];
M.~E.~Krauss and W.~Porod,
  Phys.\ Rev.\ D {\bf 92}, 055019 (2015) [arXiv:1507.04349 [hep-ph]].

\bibitem{Gluza:2015goa}
  J.~Gluza and T.~Jeli\'{n}ski,
  Phys.\ Lett.\ B {\bf 748}, 125 (2015)
  [arXiv:1504.05568 [hep-ph]].

\bibitem{KS} W.~Y.~Keung and G.~Senjanovi\'c,
  Phys.\ Rev.\ Lett.\  {\bf 50}, 1427 (1983).

\bibitem{seesaw} P. Minkowski, Phys. Lett. B {\bf 67}, 421 (1977);
R. N. Mohapatra and G. Senjanovi\'{c}, Phys. Rev. Lett. {\bf 44}, 912 (1980);
T. Yanagida, Conf.\ Proc.\ C {\bf 7902131}, 95 (1979);
M. Gell-Mann, P. Ramond and R. Slansky,  Conf. Proc. C {\bf 790927}, 315
(1979); S. Glashow, in {\em Quarks and leptons}, ed. M. L\'evy {\em et al.}, Plenum Press, NY (1980).

\bibitem{Gao:2015irw} J.~Hisano, N.~Nagata and Y.~Omura,
 Phys.\ Rev.\ D {\bf 92}, 055001 (2015) [ arXiv:1506.03931 [hep-ph]];
K.~Cheung, W.~Y.~Keung, P.~Y.~Tseng and T.~C.~Yuan,
  arXiv:1506.06064 [hep-ph];
  Y.~Gao, T.~Ghosh, K.~Sinha and J.~H.~Yu,
  Phys.\ Rev.\ D {\bf 92}, 055030 (2015) [arXiv:1506.07511 [hep-ph]];
Q.~H.~Cao, B.~Yan and D.~M.~Zhang,
  arXiv:1507.00268 [hep-ph];
T.~Abe, T.~Kitahara and M.~M.~Nojiri,
  arXiv:1507.01681 [hep-ph];
A.~E.~Faraggi and M.~Guzzi,
  arXiv:1507.07406 [hep-ph].

\bibitem{Brehmer:2015cia} J.~Brehmer, J.~Hewett, J.~Kopp, T.~Rizzo and J.~Tattersall,
  arXiv:1507.00013 [hep-ph].

\bibitem{other} J.~A.~Aguilar-Saavedra,
  arXiv:1506.06739 [hep-ph]; 
A.~Thamm, R.~Torre and A.~Wulzer,
  arXiv:1506.08688 [hep-ph]; 
A.~Carmona, A.~Delgado, M.~Quiros and J.~Santiago,
  arXiv:1507.01914 [hep-ph]; 
  Y.~Omura, K.~Tobe and K.~Tsumura,
  Phys.\ Rev.\ D {\bf 92}, no. 5, 055015 (2015)
  [arXiv:1507.05028 [hep-ph]].




\bibitem{inverse} R.~N.~Mohapatra,
  Phys.\ Rev.\ Lett.\  {\bf 56}, 561 (1986);
 R.~N.~Mohapatra and J.~W.~F.~Valle,
  Phys.\ Rev.\ D {\bf 34}, 1642 (1986).

\bibitem{Khachatryan:2015gha}
  V.~Khachatryan {\it et al.} [CMS Collaboration],
  Phys.\ Lett.\ B {\bf 748}, 144 (2015)
  [arXiv:1501.05566 [hep-ex]].


\bibitem{ATLAS2}
  G.~Aad {\it et al.} [ATLAS Collaboration],
  JHEP {\bf 1507}, 162 (2015)
  [arXiv:1506.06020 [hep-ex]].




\bibitem{HP}  M.~Cirelli, N.~Fornengo and A.~Strumia,
  Nucl.\ Phys.\ B {\bf 753}, 178 (2006) [hep-ph/0512090];
 E.~Ma and D.~Suematsu,
  Mod.\ Phys.\ Lett.\ A {\bf 24}, 583 (2009)
  [arXiv:0809.0942 [hep-ph]]; 
  Y.~Mambrini, N.~Nagata, K.~A.~Olive, J.~Quevillon and J.~Zheng,
  Phys.\ Rev.\ D {\bf 91}, 095010 (2015)
  [arXiv:1502.06929 [hep-ph]]; 
J.~Heeck and S.~Patra,
   Phys.\ Rev.\ Lett.\  {\bf 115}, 121804 (2015) [arXiv:1507.01584 [hep-ph]]. 

\bibitem{DLM}
  J.~M.~Frere, T.~Hambye and G.~Vertongen,
  JHEP {\bf 0901}, 051 (2009)
  [arXiv:0806.0841 [hep-ph]];
P.~S.~B. Dev, C.~H.~Lee and R.~N.~Mohapatra,
  Phys.\ Rev.\ D {\bf 90}, 095012 (2014)
  [arXiv:1408.2820 [hep-ph]]; 
  J. Phys.: Conf. Ser. {\bf 631}, 012007 (2015) [arXiv:1503.04970 [hep-ph]].

\bibitem{BDM} S.~Blanchet, T.~Hambye and F.~X.~Josse-Michaux,
  JHEP {\bf 1004}, 023 (2010)
  [arXiv:0912.3153 [hep-ph]];
S.~Blanchet, P.~S.~B.~Dev and R.~N.~Mohapatra,
  Phys.\ Rev.\ D {\bf 82}, 115025 (2010)
  [arXiv:1010.1471 [hep-ph]]; F.~X.~Josse-Michaux and E.~Molinaro,
  Phys.\ Rev.\ D {\bf 87}, 036007 (2013)
  [arXiv:1210.7202 [hep-ph]];
M.~Aoki, N.~Haba and R.~Takahashi,
  arXiv:1506.06946 [hep-ph]; A.~Abada, G.~Arcadi, V.~Domcke and M.~Lucente,
  arXiv:1507.06215 [hep-ph].


\bibitem{CMP} D.~Chang, R.~N.~Mohapatra and M.~K.~Parida,
  Phys.\ Rev.\ Lett.\  {\bf 52}, 1072 (1984).



\bibitem{gavela} M.~B.~Gavela, T.~Hambye, D.~Hernandez and P.~Hernandez,
  JHEP {\bf 0909}, 038 (2009)
  [arXiv:0906.1461 [hep-ph]].

 \bibitem{DP12} P.~S.~B.~Dev and A.~Pilaftsis,
  Phys.\ Rev.\ D {\bf 86}, 113001 (2012)
  [arXiv:1209.4051 [hep-ph]];
  Phys.\ Rev.\ D {\bf 87}, 053007 (2013)
  [arXiv:1212.3808 [hep-ph]].

\bibitem{Parida:2012sq}
  M.~K.~Parida and S.~Patra,
  Phys.\ Lett.\ B {\bf 718}, 1407 (2013)
  [arXiv:1211.5000 [hep-ph]];
  R.~L.~Awasthi, M.~K.~Parida and S.~Patra,
  JHEP {\bf 1308}, 122 (2013)
  [arXiv:1302.0672 [hep-ph]].

\bibitem{Abada:2014vea}
A.~Abada and M.~Lucente,
  Nucl.\ Phys.\ B {\bf 885}, 651 (2014)
  [arXiv:1401.1507 [hep-ph]].


\bibitem{AP92}  A. Pilaftsis, Z. Phys. C {\bf 55}, 275 (1992) [hep-ph/9901206].

  \bibitem{LW}
M.~Doi, T.~Kotani, H.~Nishiura, K.~Okuda and E.~Takasugi,
  Phys.\ Lett.\ B {\bf 102}, 323 (1981);
L.~Wolfenstein,
  Phys.\ Lett.\ B {\bf 107}, 77 (1981);
J.~Schechter and J.~W.~F.~Valle,
  Phys.\ Rev.\ D {\bf 23}, 1666 (1981).

\bibitem{Grimus:1998uh}
  W.~Grimus, P.~Stockinger and S.~Mohanty,
  Phys.\ Rev.\ D {\bf 59}, 013011 (1999)
  [hep-ph/9807442].

  \bibitem{DM} P.~S.~B.~Dev and R.~N.~Mohapatra,
  Phys.\ Rev.\ D {\bf 81}, 013001 (2010)
  [arXiv:0910.3924 [hep-ph]];
%
  Phys.\ Rev.\ D {\bf 82}, 035014 (2010)
  [arXiv:1003.6102 [hep-ph]].

\bibitem{Deppisch:2015qwa}
  For a review, see e.g., F.~F.~Deppisch, P.~S.~B.~Dev and A.~Pilaftsis,
  New J.\ Phys.\  {\bf 17}, 075019 (2015)
  [arXiv:1502.06541 [hep-ph]].


\bibitem{CDM} C.~Y.~Chen, P.~S.~B.~Dev and R.~N.~Mohapatra,
  Phys.\ Rev.\ D {\bf 88}, 033014 (2013)
  [arXiv:1306.2342 [hep-ph]].




\bibitem{Patra:2015bga}
  S.~Patra, F.~S.~Queiroz and W.~Rodejohann,
  arXiv:1506.03456 [hep-ph].



\bibitem{fcnc}
Y.~Zhang, H.~An, X.~Ji and R.~N.~Mohapatra,
  Nucl.\ Phys.\ B {\bf 802}, 247 (2008)
  [arXiv:0712.4218 [hep-ph]];
A.~Maiezza, M.~Nemevsek, F.~Nesti and G.~Senjanovic,
  Phys.\ Rev.\ D {\bf 82}, 055022 (2010)
  [arXiv:1005.5160 [hep-ph]];
S.~Bertolini, A.~Maiezza and F.~Nesti,
  Phys.\ Rev.\ D {\bf 89}, 095028 (2014)
  [arXiv:1403.7112 [hep-ph]].

\bibitem{Alloul:2013bka}
  A.~Alloul, N.~D.~Christensen, C.~Degrande, C.~Duhr and B.~Fuks,
  Comput.\ Phys.\ Commun.\  {\bf 185}, 2250 (2014)
  [arXiv:1310.1921 [hep-ph]].

\bibitem{Roitgrund:2014zka}
  A.~Roitgrund, G.~Eilam and S.~Bar-Shalom,
  arXiv:1401.3345 [hep-ph].

\bibitem{Alwall:2014hca}
  J.~Alwall {\it et al.},
  JHEP {\bf 1407}, 079 (2014)
  [arXiv:1405.0301 [hep-ph]].

\bibitem{Ball:2012cx}
  R.~D.~Ball {\it et al.},
  Nucl.\ Phys.\ B {\bf 867}, 244 (2013)
  [arXiv:1207.1303 [hep-ph]].

\bibitem{octet} B.~A.~Dobrescu, K.~Kong and R.~Mahbubani,
  Phys.\ Lett.\ B {\bf 670}, 119 (2008)
  [arXiv:0709.2378 [hep-ph]];
Y.~Bai and B.~A.~Dobrescu,
  JHEP {\bf 1107}, 100 (2011)
  [arXiv:1012.5814 [hep-ph]]. 


\bibitem{Langacker:1989xa}
K.~S.~Babu, K.~Fujikawa and A.~Yamada,
  Phys.\ Lett.\ B {\bf 333}, 196 (1994)
  [hep-ph/9312315].



\end{thebibliography}
\end{document}